\documentclass[epjCONF]{svjour}
\usepackage{graphicx}
\usepackage{bm}
\usepackage[varg]{txfonts} 
\usepackage[latin1]{inputenc}
\newcommand{\nuc}[2]{\ensuremath{\rm{^{#1}}#2}}         
\def\tri{{{}^3{\rm H}}}

\def\het{{{}^3{\rm He}}}

\def\be{\begin{equation}}
\def\ee{\end{equation}}
\def\bea{\begin{eqnarray*}}
\def\eea{\end{eqnarray*}}

\def\bi{\begin{itemize}}
\def\ei{\end{itemize}}
\def\n{\phantom{0}}
\def\m{\phantom{-}}

\def\jac{x}
\newcommand{\jacb}{{\bm x}}
\def\hypfi{\varphi}

\session-title{%
19$^{\textnormal{\footnotesize th}}$ International %
IUPAP Conference on Few-Body Problems in Physics%
}
\begin{document}
\title{
Proton-$\het$ elastic scattering at low energies and the ``$A_y$ Puzzle''
      }
\author{
M. Viviani\inst{1}\fnmsep\thanks{\email{michele.viviani@pi.infn.it}} 
\and %
L. Girlanda\inst{2} 
\and %
A. Kievsky\inst{1} 
\and %
L. E. Marcucci\inst{2,1}
\and %
S. Rosati\inst{2,1}
}
\institute{%
INFN, Sezione di Pisa, Largo Pontecorvo, 3, 56127 Pisa (Italy)
\and %
Phys. Dept., University of Pisa,  Largo Pontecorvo, 3, 56127 Pisa (Italy)
}
\abstract{
The Kohn variational principle and the
hyperspherical harmonic technique are applied to study
$p-\het$ elastic scattering at low energies.
Preliminary results obtained using several 
interaction models are reported. The calculations are
compared to a recent phase shift analysis performed at the Triangle University
Nuclear Laboratory and to the available experimental data.  Using a three-nucleon
interaction derived from chiral perturbation theory at 
N$^2$LO, we have found a noticeable reduction of the discrepancy
observed for the $A_y$ observable.
} 
\maketitle
%
%
%
\section{Introduction}
\label{sec:intro}

The four nucleon system has been object of intense studies
in recent years. In first place, this system is particularly interesting as a
``theoretical laboratory" to test the accuracy of our present
knowledge of the nucleon--nucleon (NN) and three nucleon (3N) interactions. 
In particular, the effect of (i) the NN P-wave and of (ii) the 3N force  are 
believed to be larger than in the $A=2$ or $3$ systems. Moreover,
it is the simplest system where the 3N interaction in  channels
of total isospin $T=3/2$ can be studied. In second place, there is a number of reactions
involving four nucleons which are of extreme importance for astrophysics, energy
production, and studies of fundamental symmetries. As an example, reactions
like $d + d \rightarrow \nuc{4}{He} + \gamma$ or $p + \nuc{3}{He} \rightarrow  
\nuc{4}{He} + \nu_{e} + e^{+}$ (the $hep$ process) play important roles in
solar models and in the theory of big-bang nucleosynthesis.

Nowadays, the four-nucleon bound state problem can be numerically solved with
good accuracy. For example, in Ref.~\cite{Kea01} the binding energies and
other properties of the $\alpha$-particle were studied using the
AV8$'$~\cite{AV18+} NN interaction; several different techniques produced
results in very close agreement with each other (at the level of less than
1\%). More recently, the same agreement has also been obtained  considering
different realistic NN+3N force models~\cite{Nogga03,Lazaus04,Viv05,Wea00}. 

In recent years, there has also been a rapid advance in solving the four nucleon
scattering problem with realistic Hamiltonians. Accurate calculations of four-body
scattering observables have been achieved in the framework of the
Faddeev-Yakubovsky (FY) equations~\cite{DF07,DF07b,DF07c}, solved in momentum
space, where the long-range Coulomb interaction is treated using the 
screening-renormalization me\-thod \cite{Alt78,DFS05}. Also
solutions of the FY equations in 
configuration space~\cite{Cie98,Lea05,Lazaus09} and several calculations using the
resonating group model \cite{HH97,PHH01,HH03,Sofia08}
were reported.  Calculations of scattering observables using the Green Function
Monte Carlo method are also underway~\cite{Wiringapc}.

In this contribution, the four-body scattering problem is solved using the 
Kohn variational method and expanding the internal part of the wave function
in terms of the hyperspherical harmonic (HH) functions (for a review,
see Ref. \cite{rep08}). Previous applications of
this method~\cite{VKR98,Vea01,Lea05} were limited to consider only
local potentials, as the Argonne $v_{18}$ (AV18)~\cite{AV18} NN potential. Recently, 
the HH method has been extended to treat also non-local
potentials, given either in coordinate- or momentum-space~\cite{Vea06,Mea09}. 
A first application of the HH method to compute $n-\tri$ elastic
observables with non-local potentials was reported in Ref.~\cite{Vea09}. Here,
we report the first results obtained for $p-\het$.

The potentials used in this paper are the I-N3LO model by Entem and
Machleidt~\cite{EM03}, with cutoff $\Lambda=500$ MeV, and the AV18. The I-N3LO
potential has been derived using an  effective field theory approach 
and the chiral perturbation theory up to next-to-next-to-next-to-leading order.
We have also performed calculations by adding to the I-N3LO potential a 
3N interaction, derived at next-to-next-to leading order
in Refs.~\cite{Eea02,N07} (I-N3LO/N-N2LO interaction model). The two free parameters in this
N-N2LO 3N potential have been chosen from the combination that reproduces
the $A=3,4$ binding energies~\cite{N07}. The development of a 3N interaction
including  N3LO contribution is still under progress~\cite{Bea07}. Finally, we
have also considered the Urbana IX (UIX) 3N potential~\cite{UIX}, which has
been used together with the AV18 potential (AV18/UIX interaction model).

The four-body studies performed so far have evidentiated several discrepancies between
the theoretical predictions and experimental data. Let us consider
first $n-\tri$ elastic scattering. Calculations based on NN 
interaction models only disagree~\cite{VKR98,Cie98,Fon99,PHH01,Lea05,DF07}
rather sizably with the measured total cross section~\cite{PBS80}, both at
zero energy and in the ``peak'' region ($E_n\approx 3.5$ MeV). Such an
observable is found to be very sensitive to the NN interaction
model~\cite{DF07}. At low energy, the discrepancy is removed by including a 3N
force fixed to reproduce the triton binding energy~\cite{VKR98,Cie98,PHH01}, but it
remains in the peak region. Interestingly, this disagreement is noticeably
reduced using the I-N3LO/N-N2LO interaction model~\cite{Vea09}. A similar
situation occurs in the differential cross section, but
definitive conclusions in this case are difficult to extract since the
experimental errors are rather large. 

In this respect, the $p-\het$ elastic scattering is more interesting  since
there exist several accurate measurements of both the unpolarized cross
section~\cite{Fam54,Mcdon64,Fisher06} and of the proton analyzing power
$A_y$~\cite{All93,Vea01,Fisher06}. The calculations performed so far (with a 
variety of NN interactions, and with the AV18/UIX model) have shown a glaring  
discrepancy between theory and experiment for
$A_y$~\cite{Fon99,Vea01,PHH01,Fisher06,DF07}. This discrepancy is very 
similar to the well known ``$A_y$  Puzzle''  in $N-d$ scattering. This is a
fairly old problem, already reported about 20 years ago~\cite{KH86,WGC88} in
the case of $n-d$ and later confirmed also in the $p-d$ case~\cite{Kie96}.
The inclusion of standard models of the 3N force has little effect on
these observables. To solve this puzzle, speculations about
the deficiency of the NN potentials in ${}^3P_j$ waves (where the
spectroscopic notation $^{2S+1}L_J$ has been adopted) have been advanced. The
situation of other $p-\het$ observables (the $\het$ analyzing power $A_{0y}$ and some
spin correlation observables as $A_{yy}$, $A_{xx}$, etc.) is less clear due to
the lack (until recently) of equally accurate measurements.

However, recently~\cite{Dan09,Dan09b} at the Triangle University National
Lab. (TUNL)  there has been a new set of accurate 
measurements (at $E_p=1.60$, $2.25$, $4$ and $5.54$ MeV) of various
spin correlation coefficients, which has allowed a phase-shift analysis (PSA).
The aim of this paper is to compare the results of the theoretical
calculations to these new data. Moreover, we want to study the effect
of including the  N-N2LO 3N force in $A_y$.

This paper is organized as follows. In Section~\ref{sec:theory}, a brief
description of the method is reported. In Section~\ref{sec:comp}, a comparison
between HH and FY calculations is shown. We have performed this comparison
for $n-\tri$ scattering using the I-N3LO potential for incident neutron energy
$E_n=3.5$ MeV. In Section~\ref{sec:res}, the theoretical calculations
are compared with the available experimental data. 
The conclusions will be given in Section~\ref{sec:conc}.

\section{The HH Technique for Scattering States}
\label{sec:theory}

In the following, we consider a $p-\nuc{3}{He}$ scattering state with
total angular momentum quantum number $JJ_z$, and parity $\pi$ (the dependence
on the wave function and other quantities on $JJ_z\pi$ will be understood in
the following).
The wave function $\Psi_{1+3}^{L S}$ describing the two particles
with incoming relative orbital angular momentum $L$ and channel spin
$S$ ($S=0, 1$) can be written as
\begin{equation}
    \Psi_{1+3}^{LS}=\Psi_C^{LS}+\Psi_A^{LS} \ ,
    \label{eq:psica}
\end{equation}
where the part $\Psi_C^{LS}$ vanishes in the limit of large inter-cluster
separations, and hence describes the system in the region where the particles
are close to each other and their mutual interactions are strong. On the other
hand, $\Psi_A^{LS}$ describes the relative motion of the two clusters in
the asymptotic regions, where the $p-\nuc{3}{He}$ interaction is
negligible (except for the long-range Coulomb interaction). In the
asymptotic region the wave functions $\Psi_{1+3}^{LS}$ reduces to
$\Psi_{A}^{LS}$, which must therefore be the appropriate asymptotic
solution of the Schr\"odinger equation. $\Psi_{A}^{LS}$ can be decomposed
as a linear combination of the following functions
\begin{eqnarray}
  \Omega_{LS}^\pm &=& D_{3+1} \sum_{l=1}^4
  \Bigl [ Y_{L}(\hat{\bf y}_l) \otimes  [ \phi_3(ijk) \otimes s_l]_{S} 
   \Bigr ]_{JJ_z} \nonumber\\
  &&\times \left ( f_L(y_l) {\frac{G_{L}(\eta,qy_l)}{q y_l}}
          \pm {\rm i} {\frac{F_L(\eta,qy_l)}{q y_l}} \right ) \ ,
  \label{eq:psiom}
\end{eqnarray}
where $y_l$ is the distance between the proton (particle $l$) and \nuc{3}{He}
(particles $ijk$), $q$ is the magnitude of the relative momentum between the
two clusters, $s_l$ the spin state of particle $l$, and $\phi_3$ is the $\het$
wave function. The total kinetic energy $T_{c.m.}$
in the center of mass (c.m.) system and the proton kinetic energy $E_p$ in the
laboratory system are
\begin{equation}
  T_{c.m.}={q^2\over 2\mu}\ , \qquad
  E_p={4\over 3} T_{c.m.}\ ,\label{eq:energy}
\end{equation}
where $\mu=(3/4)M_N$ is the reduced mass of the system ($M_N$ is the nucleon mass).
Moreover, $F_L$ and $G_L$ are the regular and irregular Coulomb function,
respectively, with $\eta=2\mu e^2/q$, and $D_{3+1}$ is a normalizing factor
defined to be 
\begin{equation}
   D_{3+1}= \sqrt{1\over 4} \sqrt{2\mu q\over\kappa_{3+1}}\ , \quad
   \kappa_{3+1}=\sqrt{3\over2}\ . \label{eq:D31}
\end{equation}
The function $f_L(y)=[1-\exp(-\beta y)]^{2 L+1}$ in Eq.~(\ref{eq:psiom})
has been introduced to regularize  $G_L$  at small $y$, and
$f_L(y) \rightarrow 1$ as $y$ is large, thus  not affecting the asymptotic
behavior of $\Psi_{1+3}^{LS}$. Note that for large values of $qy_l$,
\begin{eqnarray}
  \lefteqn{ f_R(y_l) G_{L}(\eta,qy_l)\pm {\rm i} F_L(\eta,qy_l) \rightarrow
   \qquad\qquad} &&  \nonumber \\
  && \exp\Bigl[\pm {\rm i} \bigl (q y_l-L\pi/2-\eta\ln(2qy_l)+\sigma_L\bigr ) \Bigr]\ ,
\end{eqnarray}
where $\sigma_L$ is the Coulomb phase shift.
Therefore, $\Omega_{LS}^+$  ($\Omega_{LS}^-$) describe the
asymptotic outgoing (ingoing) $p-\het$ relative motion. Finally,
\begin{equation}
  \Psi_A^{LS}= \sum_{L^\prime S^\prime}
 \bigg[\delta_{L L^\prime} \delta_{S S^\prime} \Omega_{LS}^-
  - {\cal S}_{LS,L^\prime S^\prime}(q)
     \Omega_{L^\prime S^\prime }^+ \bigg] \ ,
  \label{eq:psia}
\end{equation}
where the parameters ${\cal S}_{LS,L^\prime S^\prime}(q)$ are the $S$-matrix
elements which determine phase-shifts and (for coupled channels) mixing parameters
at the energy $T_{c.m.}$.  Of
course, the sum over $L^\prime$ and $S^\prime$ is over  all values compatible
with the given $J$ and parity $\pi$. In particular, the sum over $L^\prime$
is limited to include either even or odd values such that $(-1)^{L^\prime}=\pi$.

The \lq\lq core\rq\rq wave function $\Psi^{LS}_C$ has been here
expanded using the HH basis. For four equal mass particles, a suitable
choice of the Jacobi vectors is
\begin{eqnarray}
   \jacb_{1p}& = & \sqrt{\frac{3}{2}} 
    \left ({\bf r}_l - \frac{ {\bf r}_i+{\bf r}_j +{\bf r}_k}{3} \right )\ , \nonumber\\
   \jacb_{2p} & = & \sqrt{\frac{4}{3}}
    \left ({\bf r}_k-  \frac{ {\bf r}_i+{\bf r}_j}{2} \right )\ , \label{eq:JcbV}\\
   \jacb_{3p} & =& {\bf r}_j-{\bf r}_i\ , \nonumber
\end{eqnarray}
where $p$ specifies a given permutation corresponding to the order $i$, $j$,
$k$ and $l$ of the particles. By definition, the permutation $p=1$ is chosen
to correspond  to the order $1$, $2$, $3$ and $4$.

For a given choice of the Jacobi vectors, the hyperspherical coordinates are
given by the so-called hyperradius $\rho$, defined by
\begin{equation}
   \rho=\sqrt{\jac_{1p}^2+\jac_{2p}^2+\jac_{3p}^2}\ ,\quad ({\rm independent\
    of\ }p)\ ,
    \label{eq:rho}
\end{equation}
and by a set of angular variables which in the Zernike and
Brinkman~\cite{zerni,F83} representation are (i) the polar angles $\hat
\jac_{ip}\equiv (\theta_{ip},\phi_{ip})$  of each Jacobi vector, and (ii) the
two additional ``hyperspherical'' angles $\hypfi_{2p}$ and $\hypfi_{3p}$
defined as
\begin{equation}
    \cos\phi_{2p} = \frac{ \jac_{2p} }{\sqrt{\jac_{1p}^2+\jac_{2p}^2}}\ ,
    \quad
    \cos\phi_{3p} = \frac{ \jac_{3p} }{\sqrt{\jac_{1p}^2+\jac_{2p}^2+\jac_{3p}^2}}\ ,
     \label{eq:phi}
\end{equation}
where $\jac_{jp}$ is the modulus of the Jacobi vector $\jacb_j$. The set of angular
variables $\hat \jac_{1p}, \hat \jac_{2p}, \hat \jac_{3p}, \phi_{2p}$, and $\phi_{3p}$ is
denoted  hereafter as $\Omega_p$.  The expression of a generic HH
function is
\begin{eqnarray}
 \lefteqn{ {\cal H}^{K,\Lambda, M}_{\ell_1,\ell_2,\ell_3, L_2 ,n_2,
     n_3}(\Omega_p) =\qquad\qquad} &&  \nonumber \\
  && {\cal N}^{\ell_1,\ell_2,\ell_3}_{ n_2, n_3} 
      \left [ \Bigl ( Y_{\ell_1}(\hat \jac_{1p})
    Y_{\ell_2}(\hat \jac_{2p}) \Bigr )_{L_2}  Y_{\ell_3}(\hat \jac_{3p}) \right
    ]_{\Lambda M}  \nonumber \\
  && \times 
   \sin^{\ell_1 }\phi_{2p}    \cos^{\ell_2}\phi_{2p}
   \sin^{\ell_1+\ell_2+2n_2}\phi_{3p}
      \cos^{\ell_3}\phi_{3p}     \nonumber \\
   && \times
      P^{\ell_1+\frac{1}{2}, \ell_2+\frac{1}{2}}_{n_2}(\cos2\phi_{2p})
      P^{\ell_1+\ell_2+2n_2+2, \ell_3+\frac{1}{2}}_{n_3}(\cos2\phi_{3p})\ ,
      \label{eq:hh4P}
\end{eqnarray}
where $P^{a,b}_n$ are Jacobi polynomials and the coefficients ${\cal
N}^{\ell_1,\ell_2,\ell_3}_{ n_2, n_3}$ normalization factors. The quantity 
$K=\ell_1+\ell_2+\ell_3+2(n_2+n_3) $ is the so-called grand angular quantum
number.  The HH functions are the eigenfunctions of the hyperangular part of
the kinetic energy operator. Another important property of the HH functions is
that $\rho^K   {\cal  H}^{K,\Lambda,M}_{\ell_1,\ell_2,\ell_3, L_2 ,n_2,
n_3}(\Omega_p)$ are homogeneous polynomials of the particle coordinates of
degree $K$.

A set of antisymmetrical hyperangular--spin--isospin states of 
grand angular quantum number $K$, total orbital angular momentum $\Lambda$,
total spin $\Sigma$, and total isospin $T$  (for the given values of
total angular momentum $J$ and parity $\pi$) can be constructed as follows:
\begin{equation}
  \Psi_{\mu}^{K\Lambda\Sigma T} = \sum_{p=1}^{12}
  \Phi_\mu^{K\Lambda\Sigma T}(i,j,k,l)\ ,
  \label{eq:PSI}
\end{equation}
where the sum is over the $12$ even permutations $p\equiv i,j,k,l$, and
\begin{eqnarray}
 \lefteqn{  \Phi^{K\Lambda\Sigma T}_{\mu}(i,j;k;l)
   =\qquad\qquad} &&  \nonumber \\
  && \biggl \{
   {\cal H}^{K,\Lambda,M}_{\ell_1,\ell_2,\ell_3, L_2 ,n_2, n_3}(\Omega_p)
      \biggl [\Bigl[\bigl( s_i s_j \bigr)_{S_a}
      s_k\Bigr]_{S_b} s_l  \biggr]_{\Sigma} \biggr \}_{JJ_z}
     \nonumber \\
  && \times \biggl [\Bigl[\bigl( t_i t_j \bigr)_{T_a}
      t_k\Bigr]_{T_b} t_l  \biggr]_{TT_z}\ .
     \label{eq:PHI}
\end{eqnarray}
Here, ${\cal H}^{K,\Lambda,M}_{\ell_1,\ell_2,\ell_3, L_2 ,n_2, n_3}(\Omega_p)$ is the
HH state defined in Eq.~(\ref{eq:hh4P}), and $s_i$ ($t_i$) denotes the spin 
(isospin) function of particle $i$. The total orbital angular  momentum $\Lambda$ of
the HH function is coupled to the total spin $\Sigma$ to give the total angular
momentum $JJ_z$, whereas $\pi=(-1)^{\ell_1+\ell_2+\ell_3} $. The
quantum number $T$ specifies the total isospin of the state. The
integer index $\mu$ labels the possible choices of hyperangular, spin and
isospin quantum numbers, namely
\begin{equation}
   \mu \equiv \{ \ell_1,\ell_2,\ell_3, L_2 ,n_2, n_3, S_a,S_b, T_a,T_b
   \}\ ,\label{eq:mu}
\end{equation}
compatibles with the given values of $K$, $\Lambda$, $\Sigma$, 
$T$, $J$ and $\pi$. Another
important classification of the states is to group them in ``channels'': states
belonging to the same channel have the same values of angular
$\ell_1,\ell_2,\ell_3, L_2 ,\Lambda$, spin $S_a,S_b,\Sigma$, 
isospin $T_a,T_b,T$ quantum
numbers but different values of $n_2$, $n_3$.

Each state  $\Psi^{K\Lambda\Sigma T}_\mu$ entering 
the expansion of the four nucleon wave function must 
be antisymmetric under the exchange of any pair of particles. Consequently, it
is necessary to consider states such that
\begin{equation}
    \Phi^{K\Lambda\Sigma T}_\mu(i,j;k;l)= 
    -\Phi^{K\Lambda\Sigma T}_\mu(j,i;k;l)\ ,
     \label{eq:exij}
\end{equation}
which is fulfilled when the condition
\begin{equation} 
    \ell_3+S_a+T_a = {\rm odd}\ , \label{eq:lsa}
\end{equation}
is satisfied.

The number $M_{K\Lambda\Sigma T}$ of  antisymmetrical functions 
$\Psi^{K\Lambda\Sigma T}_\mu$
having given values of $K$, $\Lambda$, $\Sigma$, and $T$ but different
combinations of quantum numbers $\mu$ (see Eq.(\ref{eq:mu})) is in general very
large.  In addition to the degeneracy of the HH basis, the four
spins (isospins) can be coupled in different ways to $\Sigma$ ($T$). However, many
of the states $\Psi^{K\Lambda\Sigma T}_\mu$, 
$\mu=1,\ldots,M_{K\Lambda\Sigma T}$ are linearly
dependent between themselves. In the expansion of $\Psi^{LS}_C$ it is
necessary to include only the subset of linearly independent states, whose
number  is fortunately noticeably smaller than the corresponding value of
$M_{K\Lambda\Sigma T}$.

The internal part of the  wave function can be finally written as
\begin{equation}\label{eq:PSI3}
  \Psi^{LS}_C= \sum_{K\Lambda\Sigma T}\sum_{\mu} 
    u^{LS}_{K\Lambda\Sigma T\mu}(\rho)
    \Psi_{\mu}^{K\Lambda\Sigma T}\ ,
\end{equation}
where the sum is restricted only to the linearly independent states. 
We have found convenient to expand the ``hyperradial'' functions
$u^{LS}_{K\Lambda\Sigma T\mu}(\rho)$ in a 
complete set of functions, namely
\begin{equation}
     u^{LS}_{K\Lambda\Sigma T\mu}(\rho)=\sum_{m=0}^{M-1} 
      c^{LS}_{K\Lambda\Sigma T\mu,m} \; g_m(\rho)
      \ ,     \label{eq:fllag}
\end{equation}
and we have chosen 
\begin{equation}
   g_m(\rho)= 
     \sqrt{\gamma^{9}\frac{m!}{(m+8)!}}\,\,\,  
     L^{(8)}_m(\gamma\rho)\,\,{\rm e}^{-\frac{\gamma}{2}\rho} \ ,
      \label{eq:fllag2}
\end{equation}
where $L^{(8)}_l(\gamma\rho)$ are Laguerre polynomials~\cite{abra} and 
$\gamma$ is a parameter to be variationally optimized.

The main problem is the computation of the matrix elements of the Hamiltonian.
This task is considerably simplified in two steps. First, by using the
following transformation 
\begin{equation}\label{eq:arare}
  \Phi^{K\Lambda\Sigma T}_{\mu}(i,j;k;l) =
  \sum_{\mu'}  a^{K\Lambda\Sigma T}_{\mu,\mu'}(p) 
   \Phi^{K\Lambda\Sigma T}_{\mu'}(1,2;3;4)\ ,
\end{equation}
where the coefficients $a^{K\Lambda\Sigma T}_{\mu,\mu'}(p)$ have been obtained
using the techniques described in Ref.~\cite{V98}. Second,  by ``projecting'' 
the asymptotic states over a complete set of angular-spin-isospin states,
constructed in terms of the Jacobi vectors $\jacb_i$ corresponding to the
particle order $1,2,3,4$, as follows:
\begin{equation}
   \Omega_{LS}^\pm = \sum_\alpha F_\alpha^{LS\pm}(\jac_1,\jac_2,\jac_3)
   {\cal Y}_\alpha(\hat\jac_1,\hat\jac_2,\hat\jac_3)
  \ ,\label{eq:proj}
\end{equation}
where
\begin{eqnarray}
   {\cal Y}_\alpha(\hat\jac_1,\hat\jac_2,\hat\jac_3) &=& 
  \biggl\{ \biggl[\Bigl(Y_{\ell_3}(\hat\jac_3)(s_1 s_2)_{S_2}\Bigr)_{j_3}
   \Bigl(Y_{\ell_2}(\hat\jac_2)s_3 \Bigr)_{j_2}\biggr]_{J_2}
  \nonumber \\
  &&   \Bigl(Y_{\ell_1}(\hat\jac_1)s_4\Bigr)_{j_1}\biggr\}_{JJ_z}
       \Bigl[ \bigl[(t_1 t_2)_{T_2} t_3\bigr]_{T_3} t_4\Bigr]_{TT_z},
       \label{eq:proj2}
\end{eqnarray}
and $\alpha=\{\ell_1,\ell_2,\ell_3,j_1,j_2,j_3,J_2,S_2,
T_2,T_3,T\}$. Note that due to the antisymmetry of the wave function, we must
have $\ell_3+S_2+T_2=$ odd. This ``partial wave expansion'' is performed to
include all states $\alpha$ such that $\ell_i\le \ell_{\rm max}$. The
functions $F_\alpha^{LS\pm}$ can be obtained very accurately by direct
integration 
\begin{equation}
   F_\alpha^{LS\pm}(\jac_1,\jac_2,\jac_3) = 
  \int d\hat\jac_1 d\hat\jac_2 d\hat\jac_3\; \Bigl[{\cal
    Y}_\alpha(\hat\jac_1,\hat\jac_2,\hat\jac_3)\Bigr]^\dag 
  \Omega_{LS}^\pm\ ,\label{eq:proj3}
\end{equation}
using Gauss quadrature techniques.

Finally, the wave function using the transformation~(\ref{eq:arare})
and the partial wave expansion given above can be rewritten as
\begin{equation}
  \Psi_{1+3}^{LS} = \sum_\alpha {\cal
    F}_\alpha^{LS}(\jac_1,\jac_2,\jac_3)  
     {\cal Y}_\alpha(\hat\jac_1,\hat\jac_2,\hat\jac_3) 
    \ ,\label{eq:proj4}
\end{equation}
where ${\cal F}$ is a combination of Jacobi polynomials of the hyperangles,
functions $g_m(\rho)$, and
functions $F_\alpha$ coming from the asymptotic parts. Then, the matrix
elements of a two-body potential can be evaluated as following. First of all
\begin{equation}
  \langle \Psi_{1+3}^{LS}| \sum_{i<j} V(i,j) |
  \Psi_{1+3}^{L'S'}\rangle =
  6 \langle \Psi_{1+3}^{LS}| V(1,2) | \Psi_{1+3}^{L'S'}\rangle 
         \ , \label{eq:me}
\end{equation}
due to the asymmetry of the wave function. The matrix element of
$V(1,2)$ (note that $\jacb_3={\bf r}_2-{\bf r}_1$) is
easily obtained using the decomposition given in Eq.~(\ref{eq:proj4}). For
example, for a non-local potential
\begin{eqnarray}
 \lefteqn{\langle \Psi_{1+3}^{LS}| V(1,2) |
 \Psi_{1+3}^{L'S'}\rangle =\qquad\qquad} \nonumber\\
 &&= \int d^3\jacb_1 d^3\jacb_2 d^3\jacb_3 d^3\jacb_3'\;
     \Bigl(\Psi_{1+3}^{LS}(\jacb_1,\jacb_2,\jacb_3) \Bigr)^\dag
    \nonumber\\
  && \qquad\qquad \times  V(\jacb_3,\jacb_3')
      \Psi_{1+3}^{L'S'}(\jacb_1,\jacb_2,\jacb_3')
      \ . \label{eq:me1}
\end{eqnarray}
The calculation of the above integral is performed in two steps. First, the
spin-isospin-angular matrix elements 
\begin{eqnarray}
 \lefteqn{ \int d\hat\jac_1 d\hat\jac_2 d\hat\jac_3 d\hat\jac_3'\; 
     {\cal Y}_{\alpha}(\hat\jac_1,\hat\jac_2,\hat\jac_3)^\dag \;
     \qquad\qquad}&&  \nonumber \\ 
  && \qquad\qquad\qquad     
      \times V(\jacb_3,\jacb_3')\; 
      {\cal Y}_{\alpha'}(\hat\jac_1,\hat\jac_2,\hat\jac_3') 
     \nonumber\\
  &&  \quad =v^{j_3,T_3,T,T_3',T'}_{\ell_3,S_2,\ell_3',S_2'}(\jac_3,\jac_3')
   \delta_{j_3,j_3'} \delta_{j_2,j_2'} \delta_{j_1,j_1'}
   \delta_{\ell_2,\ell_2'} \delta_{\ell_1,\ell_1'}\ ,
\end{eqnarray}
are computed analytically, and, second, the integration
over the moduli of the Jacobi vectors,
\begin{eqnarray}
\lefteqn{  \int_0^\infty d\jac_1 d\jac_2 d\jac_3 d\jac_3'\; \jac_1^2\jac_2^2\jac_3^2
  \jac_3^{\prime 2} \Bigl({\cal F}^{LS}_\alpha(\jac_1,\jac_2,\jac_3)\Bigr)^*
  \qquad\qquad}\nonumber \\ 
  && \qquad\qquad \times
   v^{j_3,T_3,T,T_3',T'}_{\ell_3,S_2,\ell_3',S_2'}(\jac_3,\jac_3')\;
 {\cal F}^{L'S'}_{\alpha'}(\jac_1,\jac_2,\jac_3')\ ,
\end{eqnarray}
can easily obtained by using Gauss quadrature methods, being a
4-dimensional integral. The accuracy of the matrix
elements so calculated depends mainly on $\ell_{\rm max}$, the  maximum value
of the orbital angular momentum used to truncate the
expansion~(\ref{eq:proj}). Values $\ell_{\rm max}=5$ 
or $6$ have been found appropriate to obtain a sufficient numerical accuracy.

The $S$-matrix elements ${\cal S}_{LS,L^\prime S^\prime}(p)$ and coefficients
$c^{LS}_{K\Lambda\Sigma T\mu,m}$ occurring in the expansion of
$\Psi^{LS}_C$ are determined by making the functional
\begin{equation}
   [\overline{\cal S}_{LS,L^\prime S^\prime}(q)]=
    {\cal S}_{LS,L^\prime S^\prime}(q)
     -{1\over 2i}
        \left \langle\Psi^{L^\prime S^\prime  }_{1+3} \left |
         H-E \right |
        \Psi^{LS}_{1+3}\right \rangle
\label{eq:kohn}
\end{equation}
stationary with respect to variations in the ${\cal S}_{LS,L^\prime
S^\prime}$ and $c^{LS}_{K\Lambda\Sigma T\mu,m}$ (Kohn variational principle). 
By applying this principle, a linear set of equations
for ${\cal S}_{LS,L^\prime S^\prime}$ and $c^{LS}_{K\Lambda\Sigma T\mu,m}$
is obtained. The linear system is solved using the Lanczos algorithm.

The main difficulty of the application of the HH technique is the slow
convergence of  the basis with respect to the grand angular quantum number
$K$. This problem has been overcome by dividing the HH basis in \textit{classes}.
More details of this method can be found in Ref. \cite{Fisher06}.

\section{Comparison between HH and FY results}
\label{sec:comp}

\begin{table}
\caption[Table]{\label{table:comp}
Phase-shift and mixing angle parameters for $n-\tri$ elastic
scattering at incident neutron energy $E_n=3.5$ MeV calculated using the
I-N3LO potential. 
The values reported in the columns labeled HH have been
obtained using the HH expansion and the Kohn variational principle, whereas
those reported in the columns labeled FY by solving the FY
equations~\protect\cite{DF07b}. 
}
\begin{tabular}{l@{$\ $} c@{$\ $}c@{$\ $}|
                l@{$\ $}c@{$\ $}c}
\hline
Phase-shift & HH  & FY   & Phase-shift & HH  & FY   \\
\hline
${}^1S_0$  & $-65.66 $  & $-65.54$   & ${}^3P_0$  & $20.21$  & $20.31$   \\
\hline
${}^3S_1$  & $-58.20 $  & $-57.99$   & ${}^1P_1$  & $20.90$  & $20.74$   \\
${}^3D_1$  & $\n-0.92$  & $\n-0.91$  & ${}^3P_1$  & $40.98$  & $40.94$   \\
$\epsilon$ & $\n-0.67$  & $\n-0.72$  & $\epsilon$ & $\n9.57$ & $\n9.45$  \\
\hline
${}^1D_2$  & $\n-1.45$  & $\n-1.59$  & ${}^3P_2$  & $43.58$  & $43.98$   \\
${}^3D_2$  & $\n-0.82$  & $\n-0.84$  & ${}^3F_2$  & $\n0.05$ & $\n0.07$   \\
$\epsilon$ & $\n\m2.63$ & $\n\m2.49$ & $\epsilon$ & $\n1.14$ & $\n1.17$   \\
\hline
\end{tabular}
\end{table}

The calculated phase-shift and mixing angle parameters for $n-\tri$ elastic
scattering at $E_n=3.5$ MeV using the I-N3LO potential are reported in
Table~\ref{table:comp}. The values reported in the columns labeled HH have been
obtained using the HH expansion and the Kohn variational principle, whereas
those reported in the columns labeled FY by solving the FY
equations~\cite{DF07b}. As can be seen, there is a good overall agreement
between the results of the two calculations. Similar comparisons for $p-\het$
are currently underway.

\section{Results}
\label{sec:res}

\begin{figure}
 \includegraphics[clip,width=8cm]{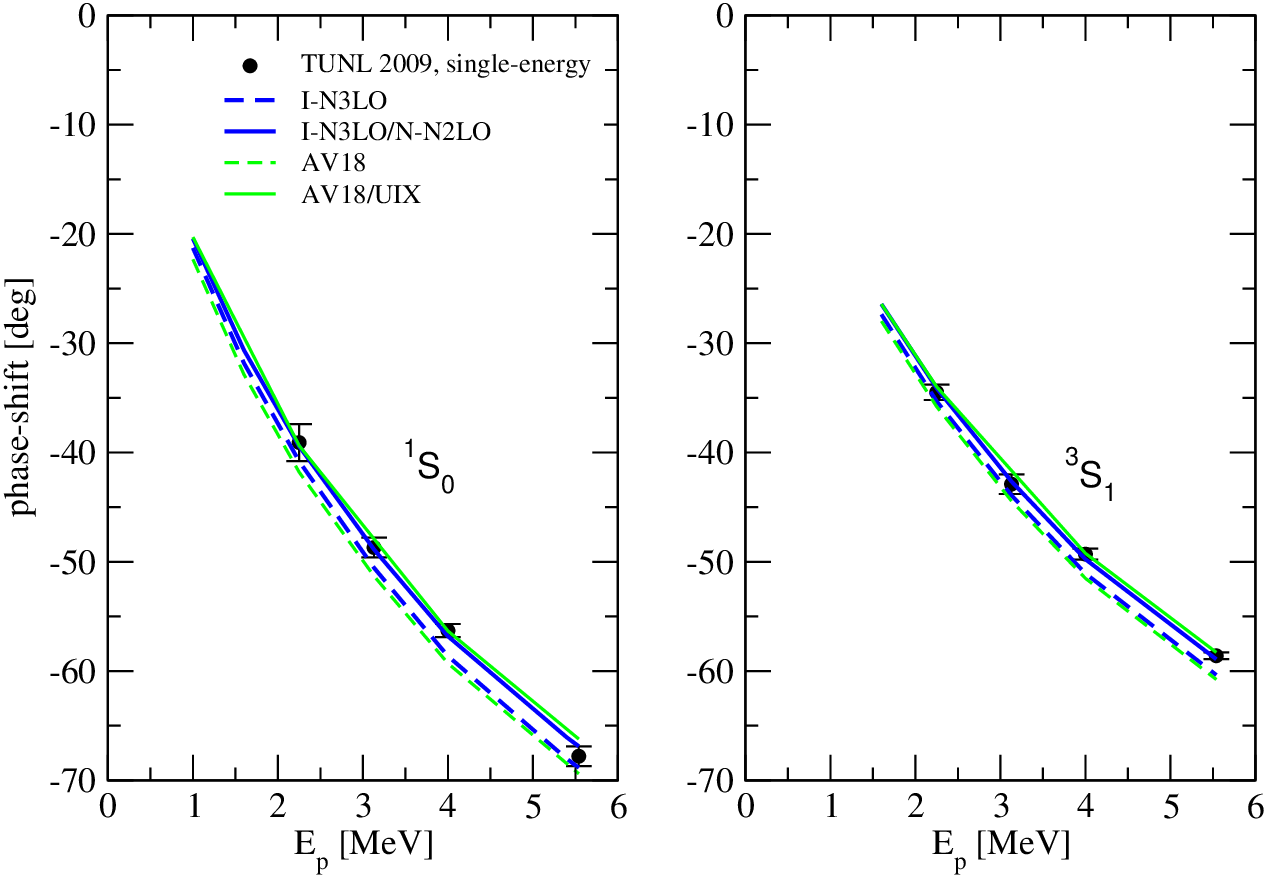}
\caption{$p-\het$ S-wave phase shifts for four potential models. The results
  of the PSA performed at TUNL have been also reported~\protect\cite{Dan09b}.}
\label{fig:psas}
\end{figure}

In this Section, we present preliminary calculations of elastic $p-\het$ 
observables. Final results with complete tests of convergence with respect
to $K$ and $\ell_{\rm maax}$ will be reported in a
forthcoming paper.

\begin{figure}
 \includegraphics[clip,width=8cm]{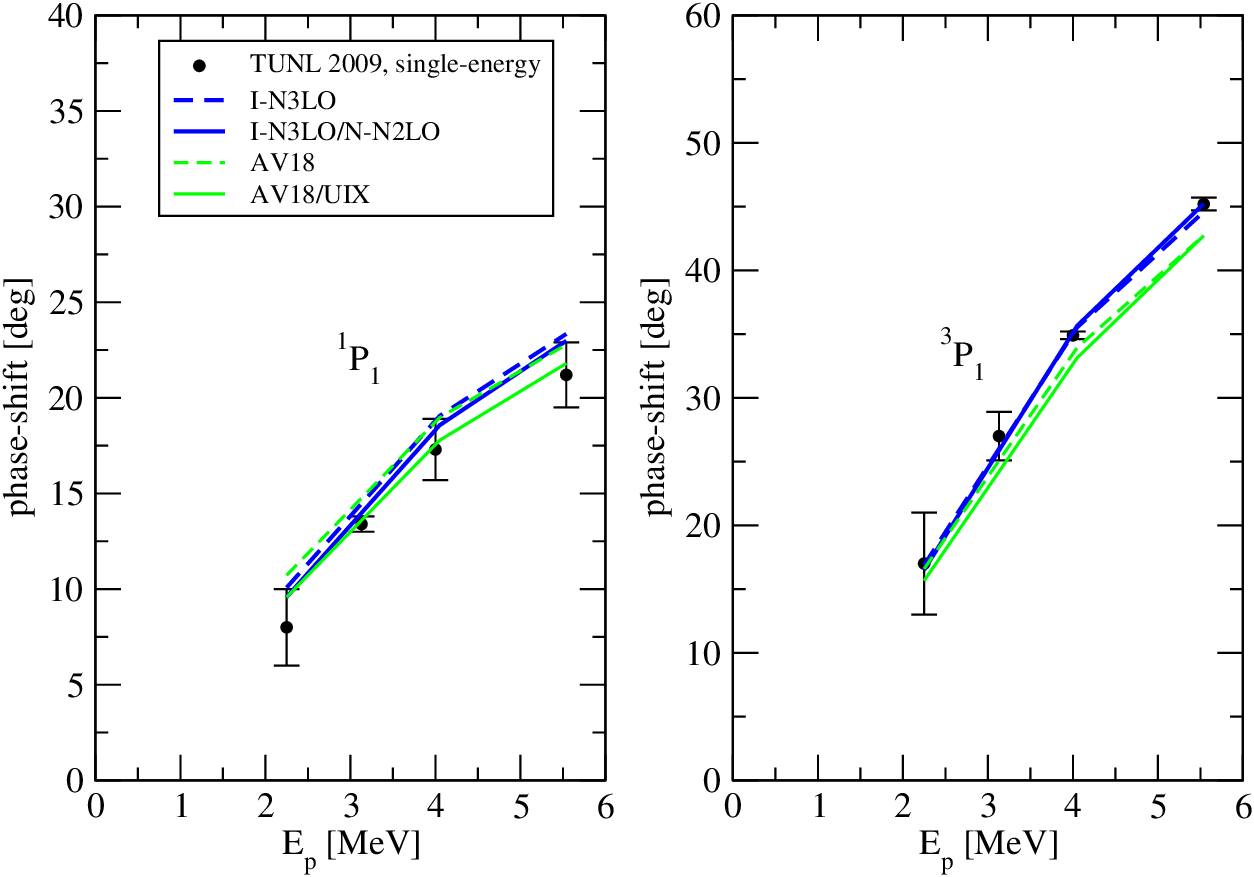}
\caption{As in Fig.~\ref{fig:psas} but for the ${}^1P_1$ and ${}^3P_1$ phase shifts.}
\label{fig:psap}
\end{figure}

\begin{figure}[b]
 \includegraphics[clip,width=8cm]{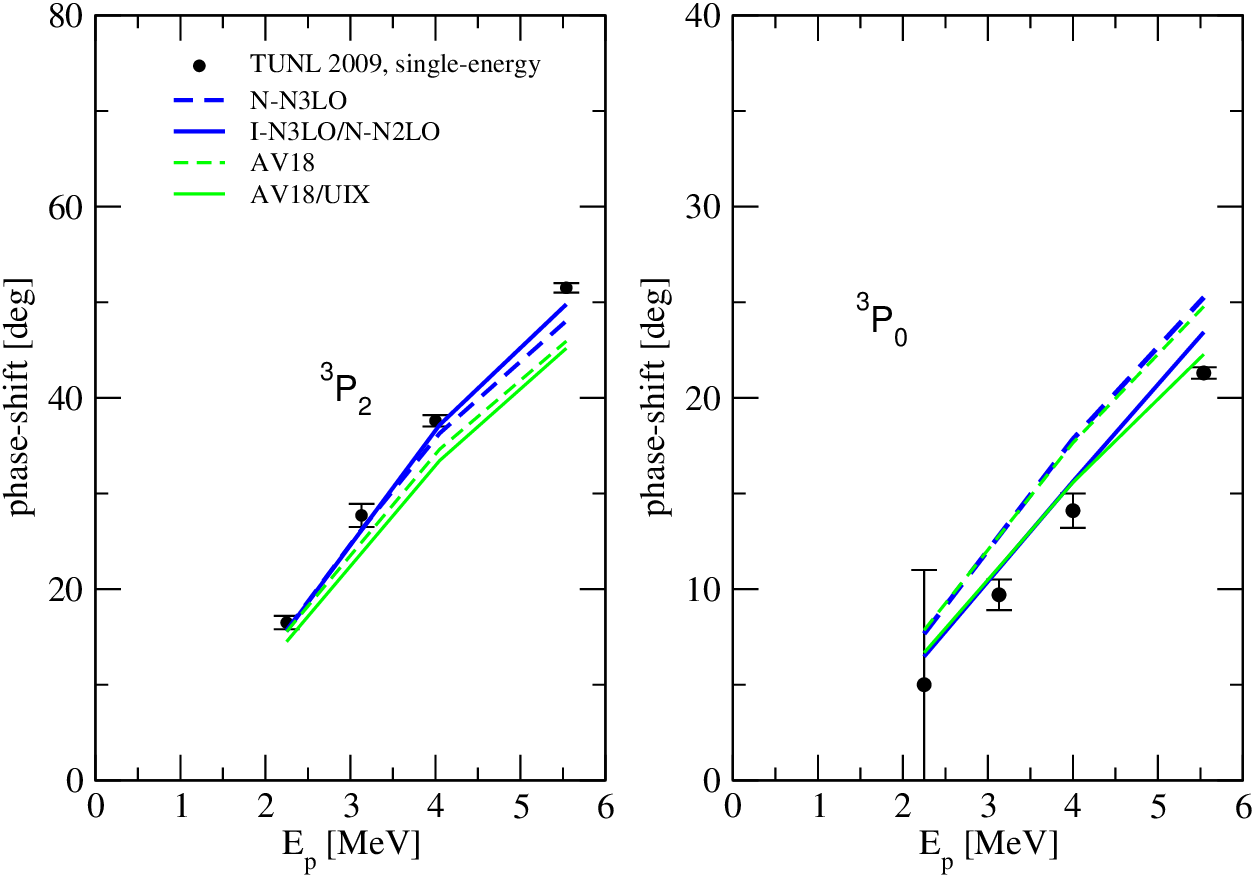}
\caption{As in Fig.~\ref{fig:psas} but for the ${}^3P_2$ and ${}^3P_0$ phase shifts.}
\label{fig:psapp}
\end{figure}

\begin{figure}[b]
 \includegraphics[clip,width=8cm]{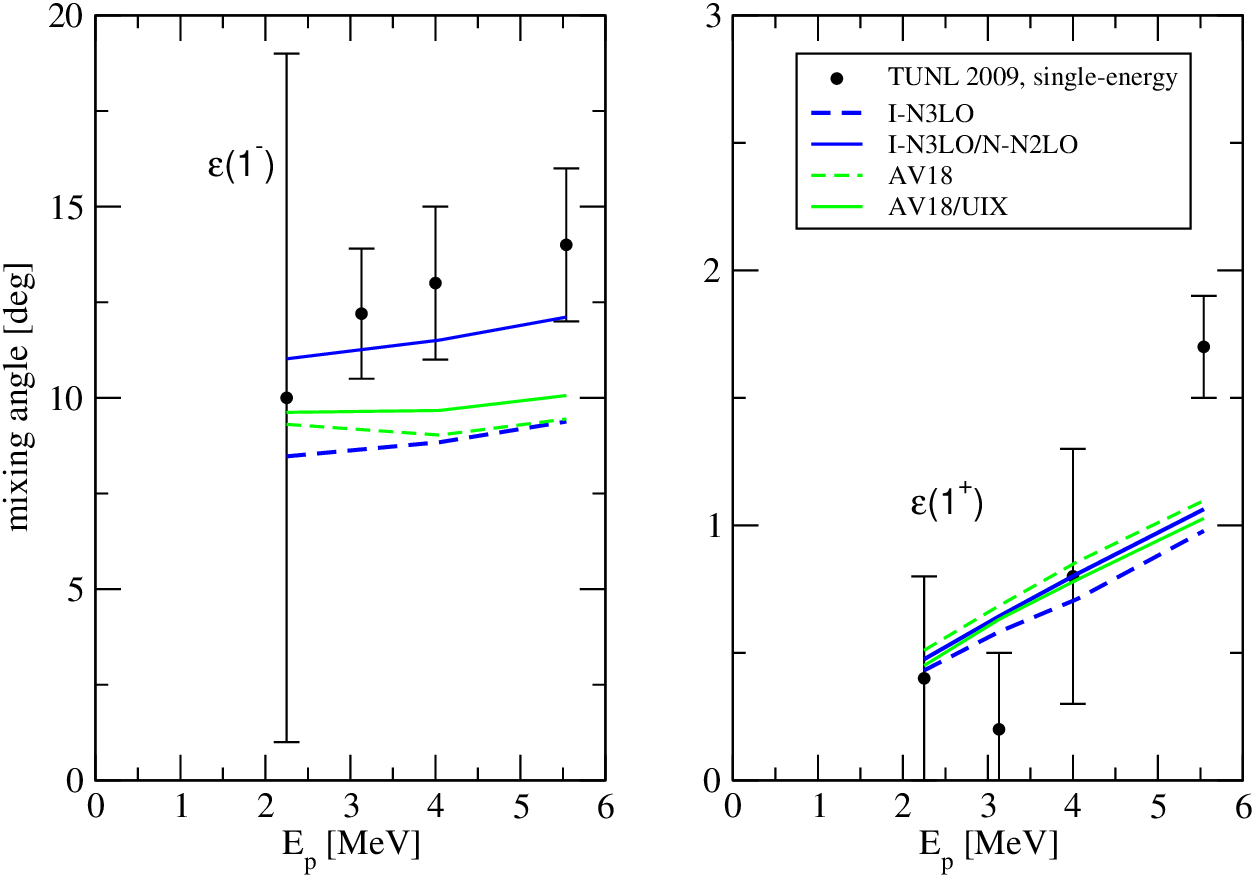}
\caption{As in Fig.~\ref{fig:psas} but for the $1^-$ and $1^+$ mixing parameters.}
\label{fig:psae}
\end{figure}

In the energy range considered here ($E_p\le 6$ MeV), the various
$p-\het$ observables are dominated by 
S-wave and P-wave phase-shifts (D-wave phase shifts give only a marginal
contribution, and more peripheral phase shifts are negligible). Let us first
concentrate on the comparison of calculated S-wave and P-wave phase
shifts with those obtained by the recent PSA~\cite{Dan09b}, shown in
Figs.~\ref{fig:psas}--\ref{fig:psae}. The ${}^1S_0$ and 
${}^3S_1$ phase shifts are reported in Fig.~\ref{fig:psas}. The results 
obtained including NN interactions only slightly overpredict
(in absolute value) the PSA values. Including the 3N force,
the calculated phase shifts agree very well with the PSA values (for
both AV18/UIX and I-N3LO/N-N2LO models). In fact, the $p-\het$
interaction in S-wave is repulsive, being dominated by the Pauli repulsion,
and the corresponding phase shifts are generally well reproduced by an
interaction model giving the correct value of the $\het$ binding energy (and
radius).

\begin{figure*}[!htbp]
 \includegraphics[clip,width=2\columnwidth]{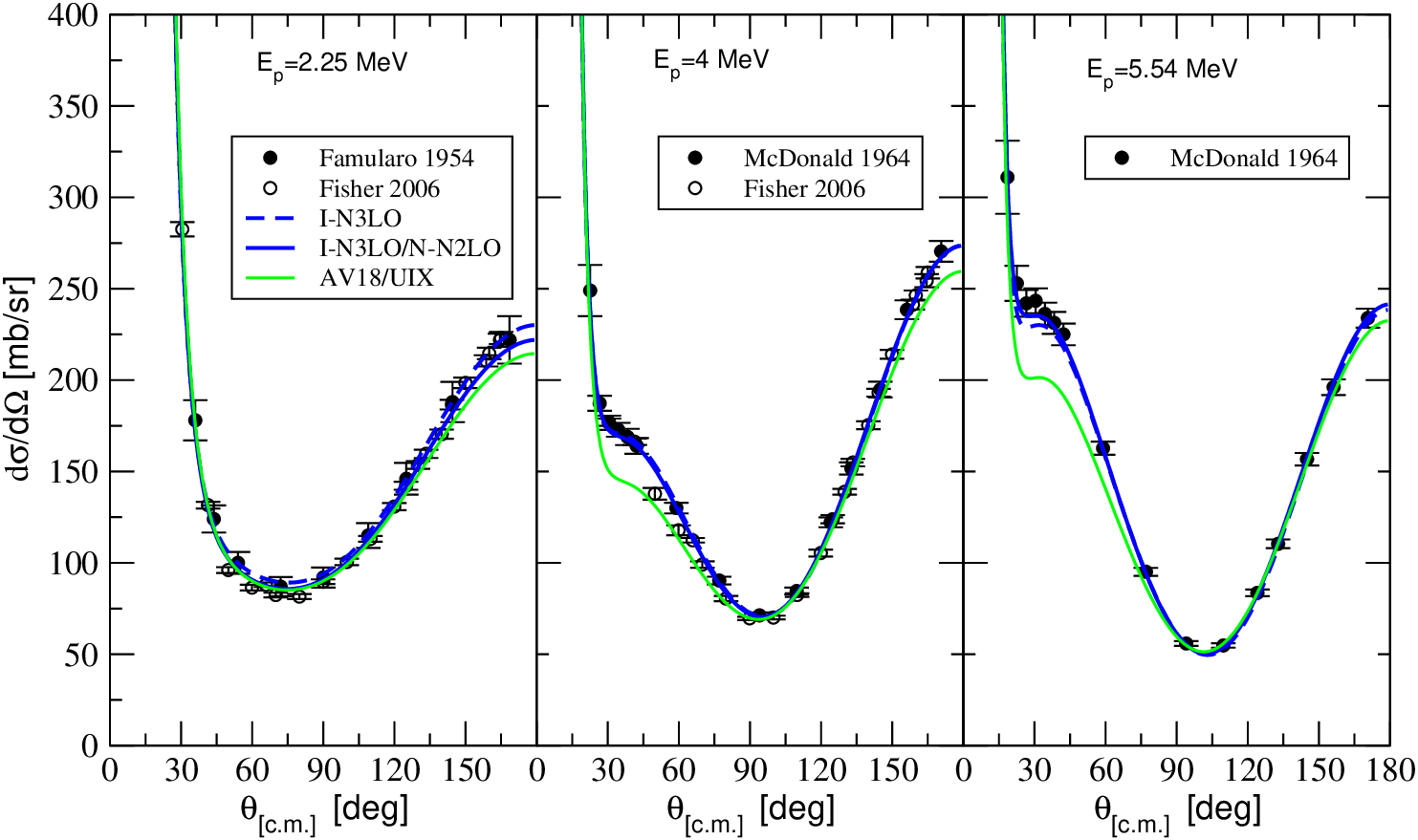}
 \caption{$p-\het$ differential cross sections calculated with the
  I-N3LO (blue dashed line), the I-N3LO/N-N2LO (blue solid line), and the
  AV18/UIX (thin green solid line) interaction  models for three
  different incident proton 
  energies. The experimental data are from
  Refs.~\protect\cite{Fam54,Mcdon64,Fisher06}. }  
\label{fig:dcs}
\end{figure*}

\begin{figure*}[!htbp]
 \includegraphics[clip,width=2\columnwidth]{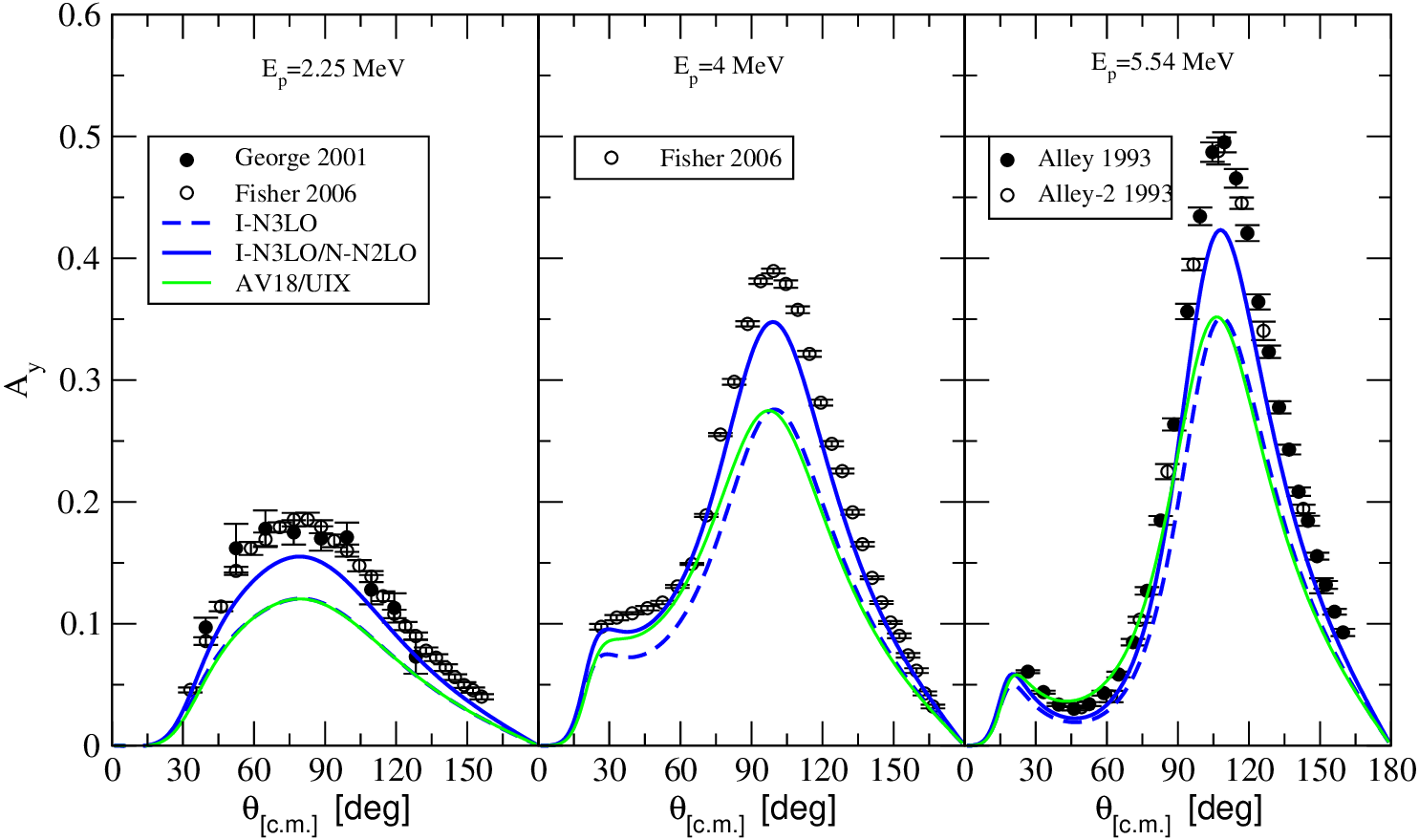}
 \caption{$p-\het$ $A_y$ observable calculated with the
  I-N3LO (blue dashed line), the I-N3LO/N-N2LO (blue solid line), and the
  AV18/UIX (thin green solid line) interaction  models for three
  different incident proton energies. The experimental data are from
  Refs.~\protect\cite{All93,Vea01,Fisher06}. }  
\label{fig:ay}
\end{figure*}

In Fig.~\ref{fig:psap}, the comparison is shown for the $J^\pi=1^-$ phase shifts.
The PSA values for the ${}^1P_1$ phase shift have
large error bars and all potential models give results consistent
with them. In general, we observe that the inclusion of the 3N interaction
reduces such a phase shift. For the ${}^3P_1$ phase shift, we
observe that the theoretical results obtained with AV18 and AV18/UIX models
disagree with the PSA values, in particular for $E_p\ge 4$ MeV. Moreover, 
the inclusion of the UIX 3N force diminishes the phase shift,
enlarging the discrepancy. On
the contrary, the results obtained with I-N3LO and I-N3LO/N-N2LO are very
close to the data. The inclusion of such a 3N force slightly increases
the ${}^3P_1$ phase shift.

A similar behavior can be observed for the ${}^3P_2$ and ${}^3P_0$ phase
shifts, reported in Fig.~\ref{fig:psapp}. The ${}^3P_2$ phase shift is well
reproduced by I-N3LO and I-N3LO/N-N2LO models, while with AV18 and AV18/UIX we
observe an underprediction. For the ${}^3P_0$ phase shift we observe
that the models including NN interaction only overpredict the PSA
values. With the inclusion of the 3N forces, 
the results come close to the PSA values. This behavior
is not unexpected, as it was already known that this
phase shift ``scales'' with the $\het$ binding energy~\cite{DF07}.

Finally, in Fig.~\ref{fig:psae} the comparison is performed for the $1^-$ and
$1^+$ mixing parameters. For $\varepsilon(1^-)$, the PSA results have
quite large errors, however it is significant that only the I-N3LO/N-N2LO
values are in the error bar. For 
$\varepsilon(1^+)$, the results obtained with all the four models are
more less consistent with the (largely scattered) PSA values.

In conclusion, with AV18, the S-wave and most of the P-wave phase shifts are
at variance with the results of the PSA. Including the UIX 3N force, only the
S-wave shifts are adjusted. With I-N3LO, the S-wave phase shifts are
overpredicted (in absolute value), but the P-wave phase shifts are much better
reproduced. Including the N-N2LO 3N interaction, the S-wave shifts become well
reproduced, while the agreement with the P-wave shifts is not spoiled. In this
case also the mixing parameters are well reproduced. We observe
therefore an overall good agreement between the results obtained with the
I-N3LO/N-N2LO model and the PSA values.

Let us now compare the results of the calculations directly with the
available experimental data. The potential models considered hereafter are the
I-N3LO, I-N3LO/N-N2LO, and AV18/UIX. The calculated $p-\het$ differential
cross sections a energies $E_p=2.25$, $4$, and $5.54$ MeV are reported in
Fig.~\ref{fig:dcs} and compared with the experimental data of
Refs.~\cite{Fam54,Mcdon64,Fisher06}.  
As can be seen, there is a good agreement between the theoretical calculations
and experimental data. In fact, this observable is sensitive to small
changes of the phase shifts only in the ``interference'' region around $\theta=30$ deg.
We note only a slight deviation of the AV18/UIX results from the data in
such a region.

More interesting is the situation for the proton vector analyzing power
$A_y$. Here, we observe a larger sensitivity to the employed
interaction model. The calculations performed using AV18/UIX
largely underpredict the experimental points, a fact already observed
before~\cite{Vea01,Fisher06}. Using I-N3LO, no significative
changes are  observed. A sizable
improvement is found by adopting the I-N3LO/N-N2LO model, as it was expected from
the discussion regarding the comparison with the PSA phase shifts. To be more
quantitative, let us consider the ratio $A_y^{\rm theor}/A_y^{\rm expt}$ at
the peak, reported in Table~\ref{table:ayp}. For the AV18, AV18/UIX, and
I-N3LO models, there is approximately a 30\% underprediction of the
peak height. The underprediction is reduced approximately by a factor
2 for the I-N3LO/N-N2LO model.  

\begin{table}[t]
\caption[Table]{\label{table:ayp}
Ratio $A_y^{\rm theor}/A_y^{\rm expt}$ at the peak for the four potential
models and three values of proton energy $E_p$.
}
\begin{tabular}{l@{$\ \ $}| c@{$\ \ $}c@{$\ \ $}c}
\hline
Pot. & $2.25$ MeV  & $4$ MeV   & $5.54$ MeV  \\
\hline
AV18          & 0.64 & 0.65 & 0.67 \\
AV18/UIX      & 0.66 & 0.70 & 0.71 \\
I-N3LO        & 0.66 & 0.70 & 0.71 \\
I-N3LO/N-N2LO & 0.87 & 0.88 & 0.86 \\
\hline
\end{tabular}
\end{table}

\begin{figure}[t]
 \includegraphics[clip,width=8cm]{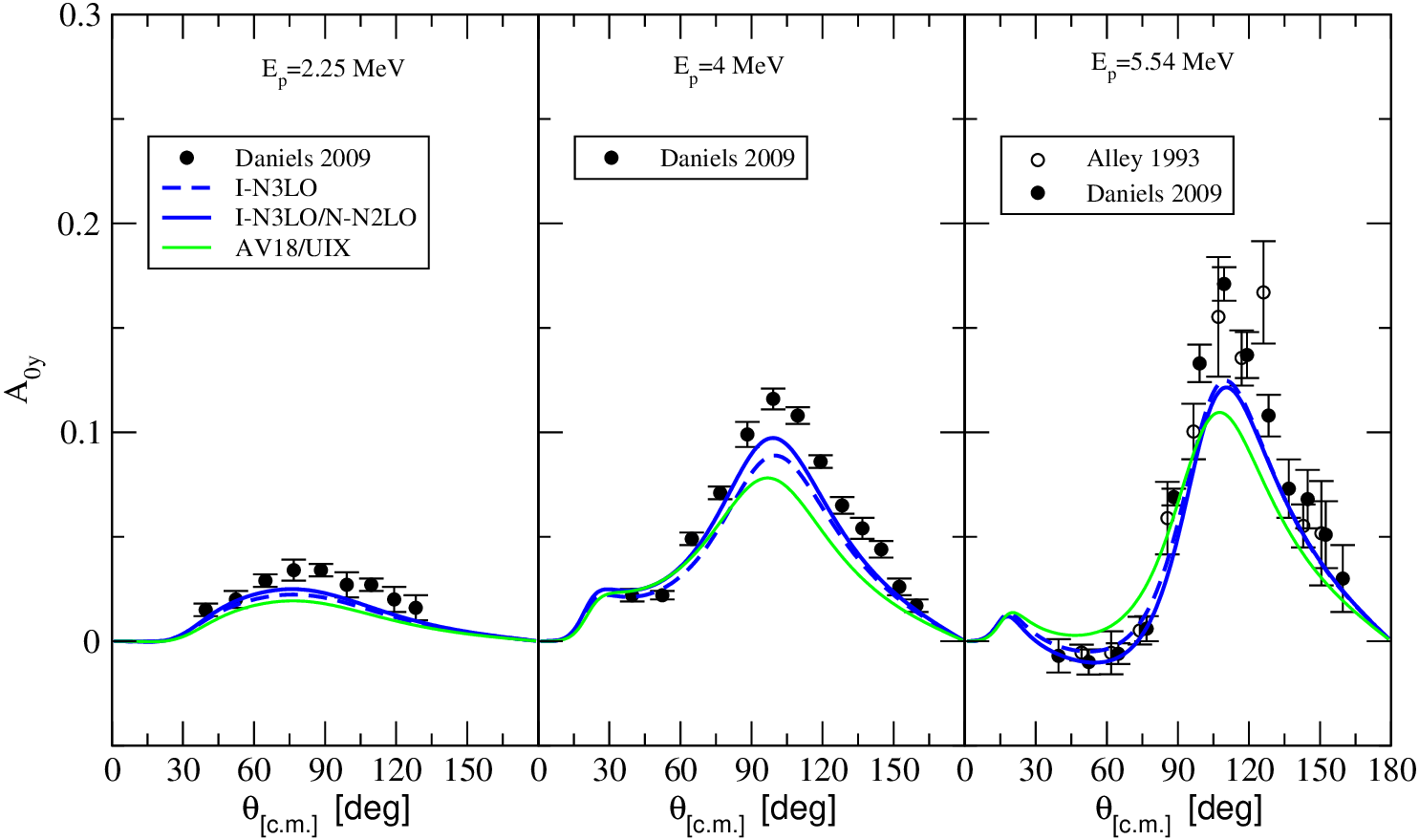}
 \caption{$p-\het$ observable $A_{0y}$ calculated with the
  I-N3LO (blue dashed line), the I-N3LO/N-N2LO (blue solid line), and the
  AV18/UIX (thin green solid line) interaction  models for three
  different incident proton energies. The experimental data
  are from Refs.~\protect\cite{All93,Dan09}. } 
\label{fig:ayh}
\end{figure}

\begin{figure}[t]
 \includegraphics[clip,width=8cm]{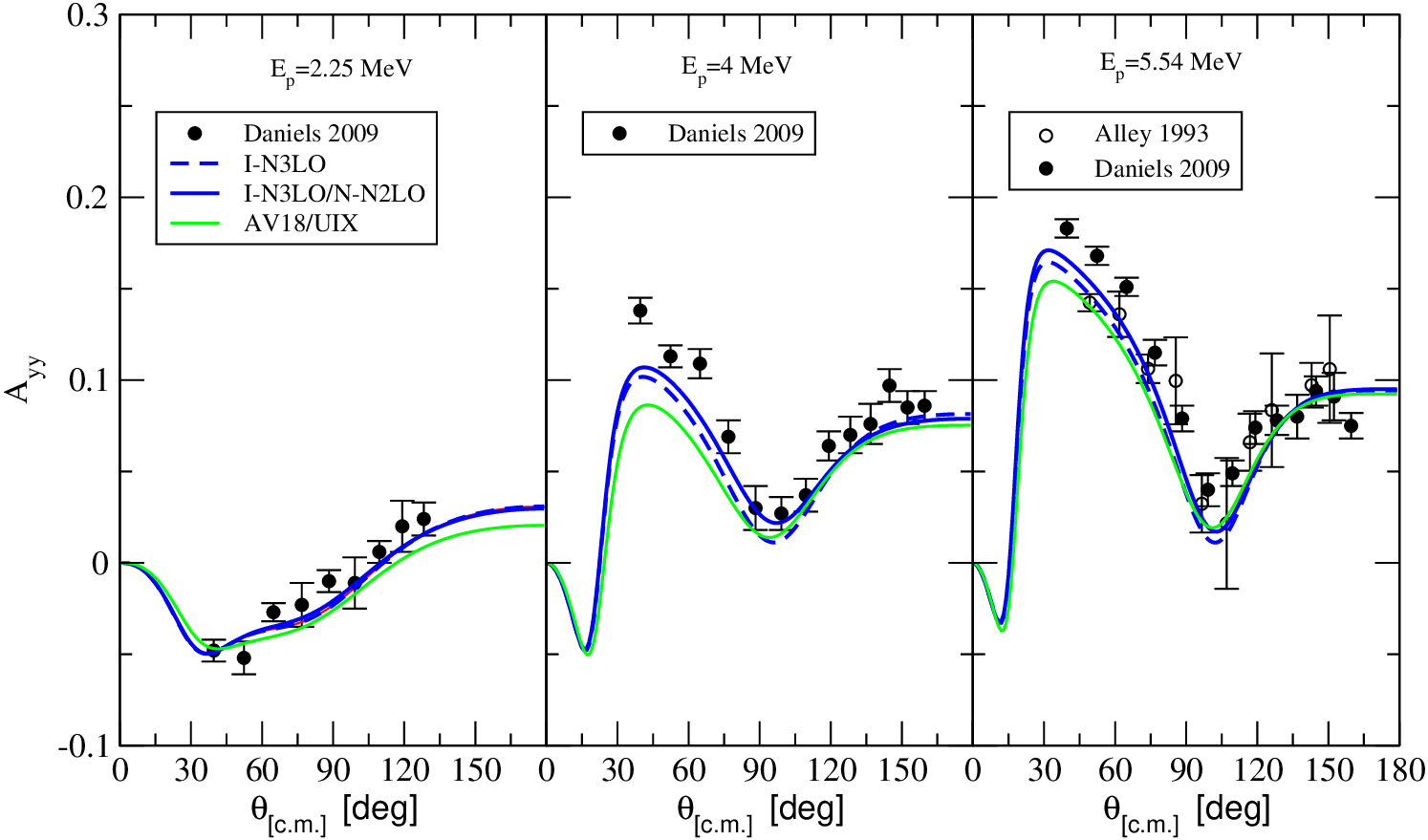}
 \caption{$p-\het$ observable $A_{yy}$ calculated with the
  I-N3LO (blue dashed line), the I-N3LO/N-N2LO (blue solid line), and the
  AV18/UIX (thin green solid line) interaction  models for three
  different incident proton 
  energies. The experimental data
  are from Refs.~\protect\cite{All93,Dan09}. } 
\label{fig:ayy}
\end{figure}

Finally, in Figs.~\ref{fig:ayh} and~\ref{fig:ayy}, we show two further
polarization observables, the $\het$ analyzing power $A_{y0}$ and the spin
correlation coefficient $A_{yy}$. These observables (and other spin correlation
coefficients) are not very sensitive to the interaction model. We observe
however, that the I-N3LO/N-N2LO model produces slightly better agreement with data.

\section{Conclusions}
\label{sec:conc}

In this paper preliminary results for $p-\het$ elastic scattering
obtained using the HH function expansion have been reported. We have
considered four interaction models. The first two are based on
phenomenological NN interactions (AV18 and AV18/UIX). The other two are derived
from a chiral effective theory (I-N3LO and I-N3LO/N-N2LO). Both NN
interactions (AV18 and I-N3LO) reproduce the NN scattering data with a
$\chi^2$ per datum very close to 1. The UIX 3N interaction has been fixed in
order to reproduce the trinucleon binding energies, whereas the N-N2LO
by fitting simultaneously both $A=3,4$ binding energies.

\begin{figure}
 \includegraphics[clip,width=.95\columnwidth]{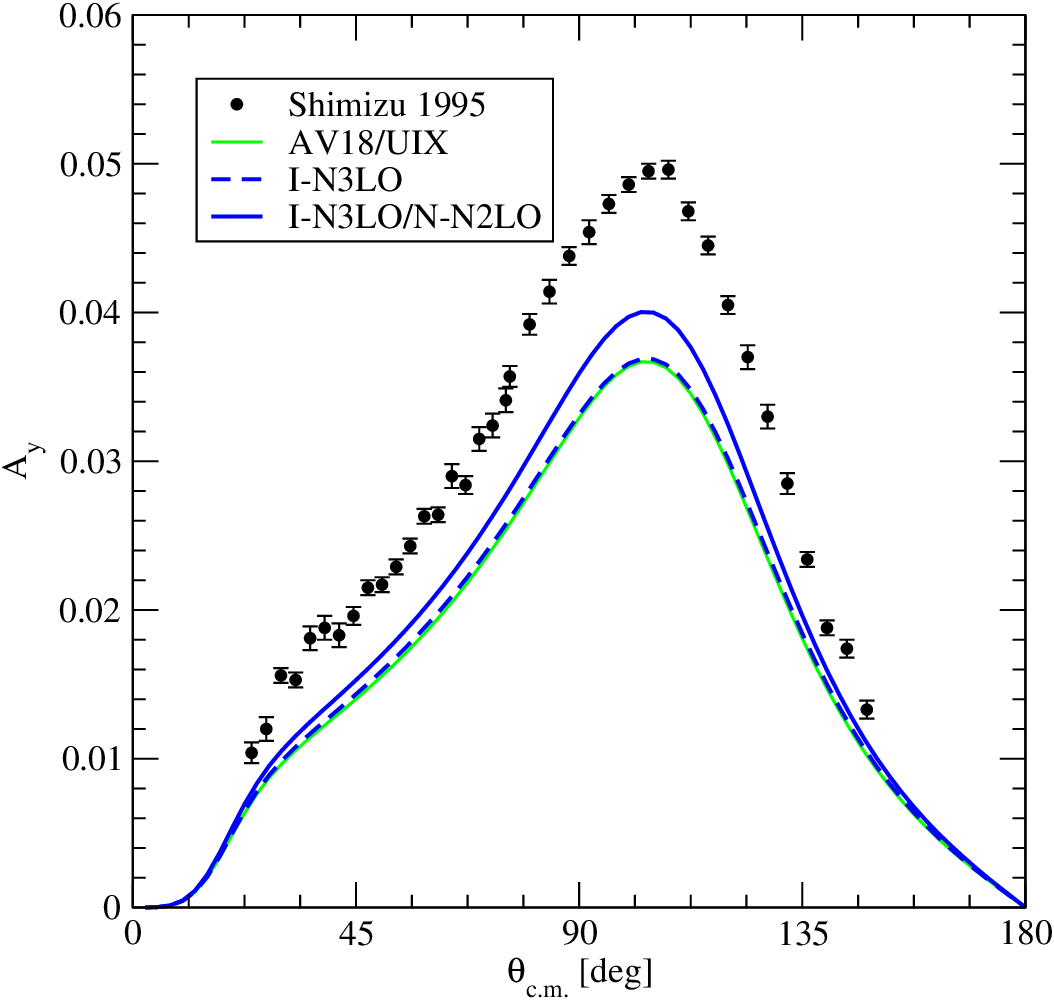}
 \caption{$p-d$ observable $A_{y}$ at $E_p=3$ MeV calculated with the
  I-N3LO (blue dashed line), I-N3LO/N-N2LO (blue solid line), and
  AV18/UIX (thin green solid line) interaction  models for $E_p=3$
  MeV. The experimental data  are from Ref.~\protect\cite{Shimi95}. } 
\label{fig:ayn}
\end{figure}

We have compared the results obtained with the four interaction models with the
available experimental data and a very recent PSA performed at
TUNL~\cite{Dan09b}. We have found that the phase shifts obtained with I-N3LO/N-N2LO are
consistent with those derived from the PSA. The direct comparison of
the I-N3LO/N-N2LO results with the experimental data has shown that there are
still some discrepancies, but the $A_y$ problem is noticeably reduced.
In fact, we observe that now the discrepancy is of the order of 15\%
at the peak, much less than what observed with other interaction models.

In this regard, it is interesting to note that in the $N-d$ case, the use of
the I-N3LO/N-N2LO model does not give such a reduction of the $N-d$
``$A_y$ Puzzle'', as can be seen for example in Fig.~\ref{fig:ayn}.
We observe in fact only a small increase of
the $A_y$ computed with the I-N3LO/N-N2LO model with respect to that
calculated, for example, using the AV18/UIX model. Note that also for $p-d$,
the calculated $A_y$ with I-N3LO and AV18/UIX are very similar. In this case the ratio
$A_y^{\rm theor}/A_y^{\rm expt}$ at the peak is
approximately 0.81 (0.74) for I-N3LO/N-N2LO (AV18/UIX). Investigations to
understand this behavior are in progress.

\medskip

{\it Acknowledgments.} We would like to thank T.V. Daniels and T.B. Clegg
for providing us with their experimental data prior to publication, and
A. Deltuva and A. Fonseca for useful discussions.


\begin{thebibliography}{99}

\bibitem{Kea01} H. Kamada {\it et al.},
               Phys. Rev. C {\bf 64}, (2001) 044001

\bibitem{AV18+} B. S. Pudliner {\it et al.},
               Phys. Rev. C {\bf 56}, (1997) 1720 

\bibitem{Nogga03} A. Nogga {\it et al.},
Phys. Rev. C {\bf 67}, (2003) 034004 

\bibitem{Lazaus04} R. Lazauskas and J. Carbonell,
Phys. Rev. C {\bf 70}, (2004) 044002 

\bibitem{Viv05} M. Viviani, A. Kievsky, and S. Rosati,
Phys. Rev. C {\bf 71}, (2005) 024006 

\bibitem{Wea00} R.B. Wiringa {\it et al.},
Phys. Rev. C {\bf 62}, (2000) 014001 

\bibitem{DF07} A. Deltuva and A. C. Fonseca, Phys. Rev. C {\bf 75},
  (2007) 014005

\bibitem{DF07b} A. Deltuva and A. C. Fonseca, Phys. Rev. Lett. {\bf 98},
  (2007) 162502 

\bibitem{DF07c} A. Deltuva and A. C. Fonseca, Phys. Rev. C {\bf 76},
  (2007)  021001  

\bibitem{Alt78} E. O. Alt, W. Sandhas, and H. Ziegelmann, Phys. Rev. C {\bf
    17},  (1978) 1981; {\it ibid.}. {\bf 21}, (1980) 1733

\bibitem{DFS05} A. Deltuva, A. C. Fonseca, and P.U. Sauer, Phys. Rev. C {\bf 71},
  (2005) 054005; {\it ibid.}, {\bf 72}, (2005) 054004

\bibitem{Cie98} F. Cieselski and J. Carbonell, Phys. Rev. C {\bf 58},
  (1998) 58; F. Cieselski, J. Carbonell, and C. Gignoux, Phys. lett. {\bf
    B447}, (1999) 199

\bibitem{Lea05} R. Lazauskas  {\it et al.}, Phys. Rev. C {\bf 71},
  (2005)  034004
  

\bibitem{Lazaus09} R. Lazauskas, Phys. Rev. C {\bf 79}, (2009) 054007 

\bibitem{HH97} H. M. Hofmann and G. M. Hale, Nucl. Phys. {\bf A613}, (1997) 69

\bibitem{PHH01} B. Pfitzinger, H. M. Hofmann, and G. M. Hale,
                Phys. Rev. C {\bf 64}, (2001) 044003 

\bibitem{HH03} H. M. Hofmann and G. M. Hale, Phys. Rev. C {\bf 68},
  (2003) 021002; Phys. Rev. C {\bf 77}, (2008) 044002 

\bibitem{Sofia08}  S. Quaglioni and P. Navr\'atil, Phys. Rev. Lett. {\bf 101},
  (2008) 092501 

\bibitem{Wiringapc} R. Wiringa, private communication

\bibitem{rep08} A. Kievsky {\it et al.}, J. Phys. G:
   Nucl. Part. Phys. {\bf 35}, (2008) 063101 

\bibitem{VKR98} M. Viviani, S. Rosati, and A. Kievsky,
                    Phys. Rev. Lett. {\bf 81}, (1998) 1580 

\bibitem{Vea01} M. Viviani {\it et al.}, 
                Phys. Rev. Lett. {\bf 86}, (2001) 3739 

\bibitem{AV18} R.B.\ Wiringa, V.G.J.\ Stoks, and R.\ Schiavilla,
               Phys.\ Rev.\ C {\bf 51}, (1995) 38 

\bibitem{Vea06} M.Viviani {\it et al.}, 
                Few-Body Syst. {\bf 39}, (2006) 159 

\bibitem{Mea09} L.E. Marcucci {\it et al.},
               Phys.\ Rev.\ C {\bf 80}, (2009) 034003 

\bibitem{Vea09} M.Viviani {\it et al.}, Few-Body Syst. {\bf 45}, (2009)
  119 

\bibitem{EM03} D.R.\ Entem and R.\ Machleidt,
               Phys.\ Rev.\ C {\bf 68}, (2003) 041001 

\bibitem{Eea02} E. Epelbaum  {\it et al.},  Phys. Rev. C {\bf 66},
                 (2002) 064001 

\bibitem{N07} P. Navr{\'a}til,  Few-Body Syst. {\bf 41}, (2007) 117 

\bibitem{Bea07} V. Bernard {\it et al.}, Phys. Rev. C {\bf 77}, (2008)
  064004 

\bibitem{UIX} B.S. Pudliner {\it et al.}, Phys. Rev. C {\bf 56},
                  (1997)  1720 

\bibitem{Fon99} A. C. Fonseca, Phys. Rev. Lett. {\bf 83}, (1999) 4021 

\bibitem{PBS80} T. W. Phillips, B. L. Berman, and J. D. Seagrave, 
                Phys. Rev. C {\bf 22}, (1980)  384 

\bibitem{Fam54} K. F. Famularo {\it et al.}, Phys. Rev. {\bf 93},
  (1954) 928 

\bibitem{Mcdon64} D. G. McDonald, W. Haberli, and L. W. Morrow,
  Phys. Rev. {\bf 133}, (1964) B1178 

\bibitem{Fisher06} B. M. Fisher {\it et al.}, Phys. Rev. C {\bf 74},
  (2006) 034001 

\bibitem{All93} M. T. Alley and L. D. Knutson, Phys. Rev. C {\bf 48},
  (1993) 1890 

\bibitem{KH86} Y. Koike and J. Haidenbauer, Nucl. Phys. {\bf A463},
               (1987) 365c 

\bibitem{WGC88} H. Witala, W. Gl\"ockle, and T. Cornelius,
                Nucl. Phys. {\bf A491}, (1988) 157 

\bibitem{Kie96} A. Kievsky  {\it et al.},
                 Nucl. Phys. {\bf A607}, (1996) 402 

\bibitem{Dan09} T.V. Daniels {\it et al.}, {\tt arXiv:1003.5860}

\bibitem{Dan09b} T.V. Daniels, private communication

\bibitem{zerni} F. Zernike and H.C. Brinkman,
    Proc. Kon. Ned. Acad. Wensch. {\bf 33}, (1935) 3 

\bibitem{F83} M. Fabre de la Ripelle,
    Ann. Phys. (N.Y.) {\bf 147}, (1983) 281 

\bibitem{abra} M. Abramowitz and I. Stegun, {\it Handbook of
    Mathematical Functions} (Dover Publications, Inc., New York, 1970)

\bibitem{V98} M. Viviani, Few-Body Syst. {\bf 25}, (1998) 177 

\bibitem{Shimi95} S. Shimizu {\it et al.}, Phys. Rev. C {\bf 52}, (1995) 1193 


\end{thebibliography}
\end{document}